\documentclass{iau} 
\usepackage{graphicx}

\title[Dynamical modelling of galactic disc outskirts] %% give here short title %%
{Dynamical modelling of galactic disc outskirts}

\author[E. Athanassoula]   %% give here short author list %%
{E. Athanassoula}

\affiliation{Aix Marseille Univ, CNRS, LAM, Laboratoire d'Astrophysique de Marseille, Marseille, France \\ email: {\tt lia@lam.fr}}
 
\pubyear{2016}
\volume{321}  %% insert here IAU Symposium No.
\jname{Formation and evolution of galaxy outskirts}
\editors{A. Gil de Paz, J.C. Lee \& J.H. Knapen, eds.}
\begin{document}

\maketitle

\begin{abstract}
I review briefly some dynamical models of structures in the outer
parts of disc galaxies, including models of polar rings, tidal
tails and bridges. I then discuss the density distribution in the
outer parts of discs. For this, I compare observations to results
of a model in which the disc galaxy is in fact the remnant of a major
merger, and find good agreement. This comparison includes radial
profiles of the projected surface density and of stellar age, as well
as time
evolution of the break radius and of the inner and outer disc scale
lengths. I also compare the radial projected surface density profiles
of dynamically motivated mono-age populations and find that, compared
to older populations, younger ones have flatter density profiles in
the inner region and steeper in the outer one. The 
break radius, however, does not vary with stellar age, again in good
agreement with observations.  

\end{abstract}

\firstsection % if your document starts with a section,
\section{Introduction}

What distinguishes the outskirts from the inner parts  of galactic
discs, to the extent that a specific IAU symposium should be devoted
to their study? Why and how is the modelling of galactic disc outskirts
different from that of their inner parts? Several points come
immediately to mind. 

Compared to the inner parts, the outer parts of galactic discs:
\begin{itemize}
\item
are more exposed to accretion, to interactions and to minor mergers.

\item 
have dynamics which are more dominated by the (spheroidal) halo than by the (thin) disc.

\item
have stronger asymmetries (m=1 features) and multi-arm (high-m) spirals, but have no bars.

\item
formed more recently that the inner parts, given the generally admitted inside out disc formation.
\end{itemize}

\vspace{0.5cm}
As a result:
\begin{itemize}
\item 
the outer parts of discs are well suited for studies of galaxy environment, accretion, interactions and mergers, i.e. for many drivers of galactic evolution (secular or not).
\item 
their study can lead to invaluable constraints on the mass of dark matter haloes, its distribution and shape.
\item 
their outer parts allow us to study disc assembly using nearby galaxies, rather than galaxies at higher redshifts, and thus at much higher resolution and sensitivity.
\end{itemize}

Here I will review briefly several attempts to model structures in the
outskirts of disc galaxies. These will include polar rings
(Sect. \ref{sec:polar-rings}), tidal tails and bridges
(Sect. \ref{sec:bt}) and type II outer discs
(Sect. \ref{sec:outer-discs}). Warps and type III profiles were included in the oral version of this talk, but are not included here due to strict time and space limitations. 

\section{Polar rings}
\label{sec:polar-rings}

Polar rings\footnote{For more detailed reviews of this subject see
\cite{Athanassoula.Bosma.85}, \cite{Schweizer.96} and \cite{Sparke.04}.} are rings of gas, dust and stars rotating in a plane
perpendicular to the main disc, in a polar configuration. They are
relatively rare objects  in S0 galaxies and even rarer in spirals
(\cite{Athanassoula.Bosma.85}, \cite{Whitmore.etal.90}, \cite{Reshetnikov.FO.11}). 
NGC 4650A is a prototype polar ring
galaxy. 

The
formation of two discs at right angles seems a most unlikely thing to
happen in situ, so that one can safely conclude that polar rings are due to some
external event such as the accretion of an external satellite galaxy,
or perhaps a mass transfer during an encounter. A number of models
around this basic scheme have been proposed (\cite{Bekki.97},
\cite{Reshetnikov.S.97}, \cite{TremaineYu.00}, \cite{BournaudCombes.03}, etc.). 
More recently \cite{Maccio.Moore.Stadel.06}, \cite{Brook.etal.08} and \cite{Snaith.etal.12}, using
cosmological simulations and their zoom re-simulations, showed that polar 
rings can be an extreme case of accretion of gas whose angular
momentum is near-perpendicular to that of the central galaxy.

Polar rings offer a unique opportunity for probing galactic potentials far
from the galactic centre and in a polar plane. They, therefore, can
be used to get information on the shape of the dark matter halo
(e.g. \cite{Whitmore.etal.87}, \cite{Sackett.Sparke.90},
\cite{Sackett.etal.94}, \cite{Combes.Arnaboldi.96}, Iodice et
al. 2003, \cite{Khoperskov.etal.14}). This last paper provides also a
statistics of the values of the halo minor-to-major axis ratio ({\it{c/a}}) obtained for
polar ring galaxies. Their histogram shows a strong clear peak near
{\it c/a} = 1, i.e. a
strong preference for near-spherical shapes. The total distribution,
however, is bimodal, with a second, albeit much less important, maximum around
{\it c/a} = 0.4. The physical significance of this second peak is unclear, but
Khoperskov and his collaborators underline that the {\it{c/a}} value is in fact a function
of distance from the centre, with the central parts (where the
gravitational influence of the disc is most important) being flatter than
the outer parts, which can be more spherical or even
prolate. They, furthermore, stress the existence of limitations arising
from model assumptions and observational difficulties. 
 
%\section{Warps}

\section{Tidal bridges and tails}
\label{sec:bt}

Tidal bridges and tails\footnote{For more detailed reviews of this subject see
\cite{Barnes.Hernquist.92a}, \cite{Barnes.96}, \cite{Schweizer.96},  
\cite{Struck.99}, etc.} are very spectacular structures which may form during
close encounters of two galaxies. Bridges are relatively narrow
and link the two galaxies, while their shape -- in particularly in their
central part -- can appear not far from
linear. Each of the two discs forms also a long curving tail which
emanates from the part of the disc which is roughly opposite to the
other galaxy. This tail can reach very large distances from its parent
galaxy. Prototypes of such structures are NGC 4038/4039 (better known as the
Antennae) and NGC 4676 (better known as the Mice), which have been
extensively modelled. \cite{TT72} deliberately used very simple test
particle simulations in which each of the two galaxies is a massive point
surrounded by massless test particles which just follow the potential.
With these they performed 
ground breaking work in this field, and established that gravity on
its own is sufficient to explain such features.
They also showed that the results of direct
encounters (where the sense of rotation of the companion is the same 
as that of the particles in the disc) are much more spectacular than
the results of retrograde encounters (where the sense of rotation of
the companion is opposite to that of the particles in the disc). 

These simulations were followed by more realistic ones, in which the gravity of all
components (disc, halo and bulge) is fully taken into account and
calculated at each simulation step. These of course necessitate much more
computer time.  
The necessary computer time was even more substantially increased when
gas, as well as its physics (star formation, feedback and 
cooling) were included in the simulations of interacting and merging
systems. Nevertheless, the introduction of self-consistency and of gas
did not alter much the large-scale dynamics and corresponding
formation of bridges and tails, so that the main results of \cite{TT72} 
were not substantially affected (\cite{Barnes.Hernquist.96}). It
allowed, however, studies which could not have been
done with the previous techniques. 

\cite{Dubinski.MH.96} suggest that the length of the tidal tails
can be used to estimate the masses of dark matter haloes in disc galaxies. 
Indeed, they found that for more massive and more extended dark haloes, 
the resulting tidal tails become shorter, less
massive, and less striking, even under the most favourable conditions
for producing these features. Thus the large halo-to-disc mass ratio
expected in $\Lambda$CDM cosmologies seems at odds with the 
observed tidal tails. This conclusion, however, was based on a rather
restricted number of progenitor models (but see also
\cite{Dubinski.MH.99} for a more extended set of
models). \cite{Springel.White.99} using a set of progenitor models based
on the \cite{Mo.Mao.White.98} analytical models, demonstrate that even
the discs which are embedded in 
very massive dark matter haloes could well develop long tidal tails,
provided the halo spin parameter is large enough. They thus argued
that halo-to-disc mass ratio is not a good indicator of the ability to
produce tails, and cast doubts as to whether tidal tails could
be expected to constrain the values of the cosmological parameters.  

\cite{Bournaud.Duc.Masset.03} and \cite{Duc.Bournaud.Masset.04}
revisited this question, now using the 
existence of massive gaseous accumulations near the tip of the
tidal tails, and showed that this requires that the dark matter haloes
extend at least ten times further than the stellar discs. These gaseous
accumulations could in fact be progenitors of tidal dwarf galaxies
(see also \cite{Barnes.Hernquist.92b}). 

A poster presented at this meeting 
%\citep
(\cite{RomeroGomez.Athanassoula.16}) discussed whether it is
possible to use manifolds to understand the dynamics underlying the
formation of bridges and tails. They considered
the galaxies as point masses and also used the restricted three-body
approach. They studied the 
positions and stability of the Lagrangian points in this system and
found that, at least for such a simplified model, manifold shapes
allow for bridges and 
tails with realistic shapes and properties.  

\section{Surface density radial profiles: Breaks and outer discs}
\label{sec:outer-discs}

Radial luminosity profiles are available for a
large number of disc galaxies, both spirals and S0s, and in many
wavelengths. The inner parts of these profiles  
are dominated by the light from the classical bulge and/or
from the discy pseudobulge\footnote{See \cite{Kormendy.Kennicutt.04}
and \cite{Athanassoula.05} for a discussion of the different types of bulges.}.
Beyond that, at radii where the disc  
is the main contributor to the luminosity, the radial profile shows 
single, double, or even multiple 
segments of exponential profiles. Single exponential profiles are
called type I, while the double ones are known either 
as type II (downbending), or type III (upbending, or
anti-truncating), depending on whether the decrease of the luminosity
with radius  in
the outer parts is more, or less, steep than that of the inner
parts (e.g. \cite{Freeman.70}, \cite{Pohlen.DLA.02}, \cite{Pohlen.Trujillo.06},
\cite{Erwin.PB.08}, \cite{Gutierrez.EAB.11}, 
\cite{Munoz-Mateos.S4G.13}, \cite{Laine.P.14}, 
\cite{Kim.P.14}). The radius at which
the inner and outer 
exponential profiles intersect is known as the break radius.      

Here I will discuss the formation and evolution of such profiles in the
context of models in which a disc galaxy forms from a major merger. Such
models are described in some detail in 
\cite{Athanassoula.RPL.16a}, which I will hereafter refer to as A16a, while 
their salient features are summarised in the Appendix. 
Peschken, Athanassoula \& Rodionov (2016, hereafter P16) and 
Athanassoula, Peschken \& Rodionov (2016b, hereafter A16b) 
analysed about  100 such simulations. The
results of Sect.~\ref{subsec:typeIIa} and \ref{subsec:typeIIb} come from
these works and should be considered as preliminary. They are also
restricted to type II profiles. No type I model was found in these
models and for
types III, the reader is referred to A16b, and to the references
therein. 

\subsection{Profiles of type II: General considerations and trends
  with age and time}
\label{subsec:typeIIa}

\begin{figure}[t]
% \vspace*{-2.0 cm}
\begin{center}
 \includegraphics[width=3.9in]{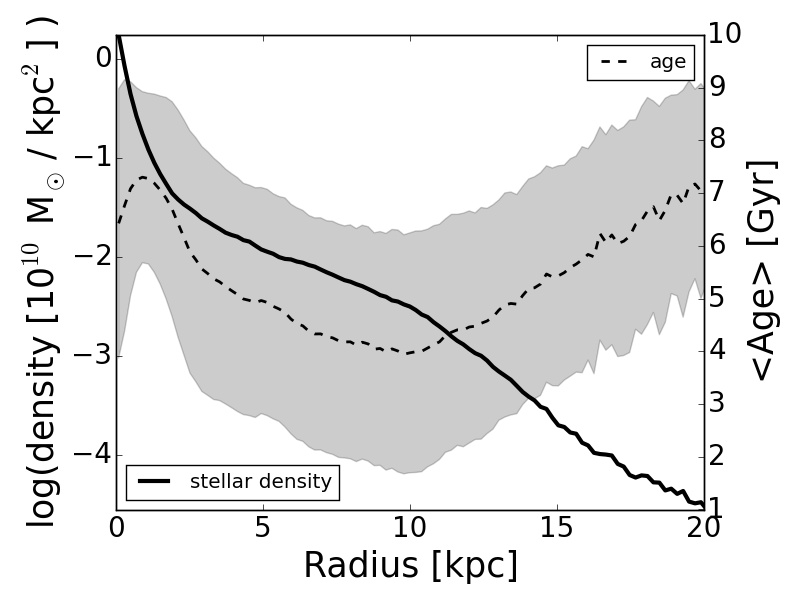} 
% \vspace*{-1.0 cm}
 \caption{Stellar projected surface density (solid line) and mean age
   (dashed line) both as a function of distance from the centre from a
   simulation of a major merger of two disc protogalaxies, each in
   its halo of dark matter and hot gas.}
   \label{fig:typeII}
\end{center}
\end{figure}

Profiles of type II are the most common ones both in observations
of field galaxies (e.g. \cite{Pohlen.Trujillo.06}, \cite{Erwin.PB.08}, 
\cite{Laine.P.14}) and in simulations, and thus have been more
extensively studied. Fig.~\ref{fig:typeII} shows a clear example of
such a profile (solid line) from one of the models described 
in A16a. It also shows
the mean age of the stars as a function of radius (dashed
line). This has a U shape with a 
clear minimum at a location roughly coinciding with that of the
break. These characteristic profile shapes for the density/luminosity
and the age, as well as the coincidence of the break
radius with the location of the minimum of the age profile have been
often observed in disc galaxies (e.g. \cite{Bakos.TP.08}, and
\cite{Azzollini.TB.08a}, but see also e.g.
\cite{Roediger.CSM.12}, and \cite{Ruiz-Lara.P.16} for a different
point of view), as well as in simulations 
with isolated disc galaxy initial conditions (\cite{Roskar.DSQ.08}). 

The simulations of A16b show also that the radial surface density profile of
the disc component evolves secularly, and in particular that the break
radius increases considerably 
with time in good agreement with the observations of
\cite{Perez.04}, of \cite{Azzollini.Trujillo.Beckman.08b} and
\cite{Munoz-Mateos.BG.11}, as well as with a general inside-out 
growth of the disc component. 

It is also possible to get information on the evolution using fossil
records, i.e. examining separately populations of different ages. 
A16b define their mono-age populations in two different
ways. In the first, they follow \cite{Martig.Minchev.Flynn.14} and
divide the simulation time (from 0 to 10 Gyr) in age brackets of equal
duration. Since the evolution is not very rapid A16b use only five age 
brackets, calculate the radial profiles and measure  the 
breaks and the inner and outer disc scale lengths for each age bracket 
separately. In the second way, A16b divide the time into brackets of unequal
lengths whose limits are set by landmark times, i.e. times that
strongly mark the evolution, like the merging time and the time after
which the thin disc starts forming
(see A16a). In both cases they find that the break
radii are, to within the measuring errors, independent of the age
bracket chosen, while inner (outer) disc scale length decreases
(increases) as the age of the population increases. This is in good
agreement with observations as shown e.g. for NGC 4244 by  
\cite{deJong.P.07}, for NGC 7793 by \cite{Radburn-Smith.RDD.12}
and for our Galaxy e.g. by \cite{Sale.P.10}, and by references therein.

To test an  extreme population difference, A16b also compared the
break radius of the  
star forming gas to that of the total stellar component at {\it{t}}=10 Gyr for their
sample of simulations. They 
find them to be equal to within the measuring errors, again 
in good agreement with observations. Indeed, this is what was found by 
\cite{Munoz-Mateos.P.14}, who compared the radial projected density
profile of the molecular gas to that of the old
stellar population of NGC 5985, as obtained by the Spitzer Survey of Stellar
Structure in Galaxies (S$^4$G, \cite{Sheth.S4G.10}).

\subsection{Profiles of type II: Trends with angular momentum}
\label{subsec:typeIIb}
 
\begin{figure}[t]
% \vspace*{-2.0 cm}
\begin{center}
 \includegraphics[width=5.2in]{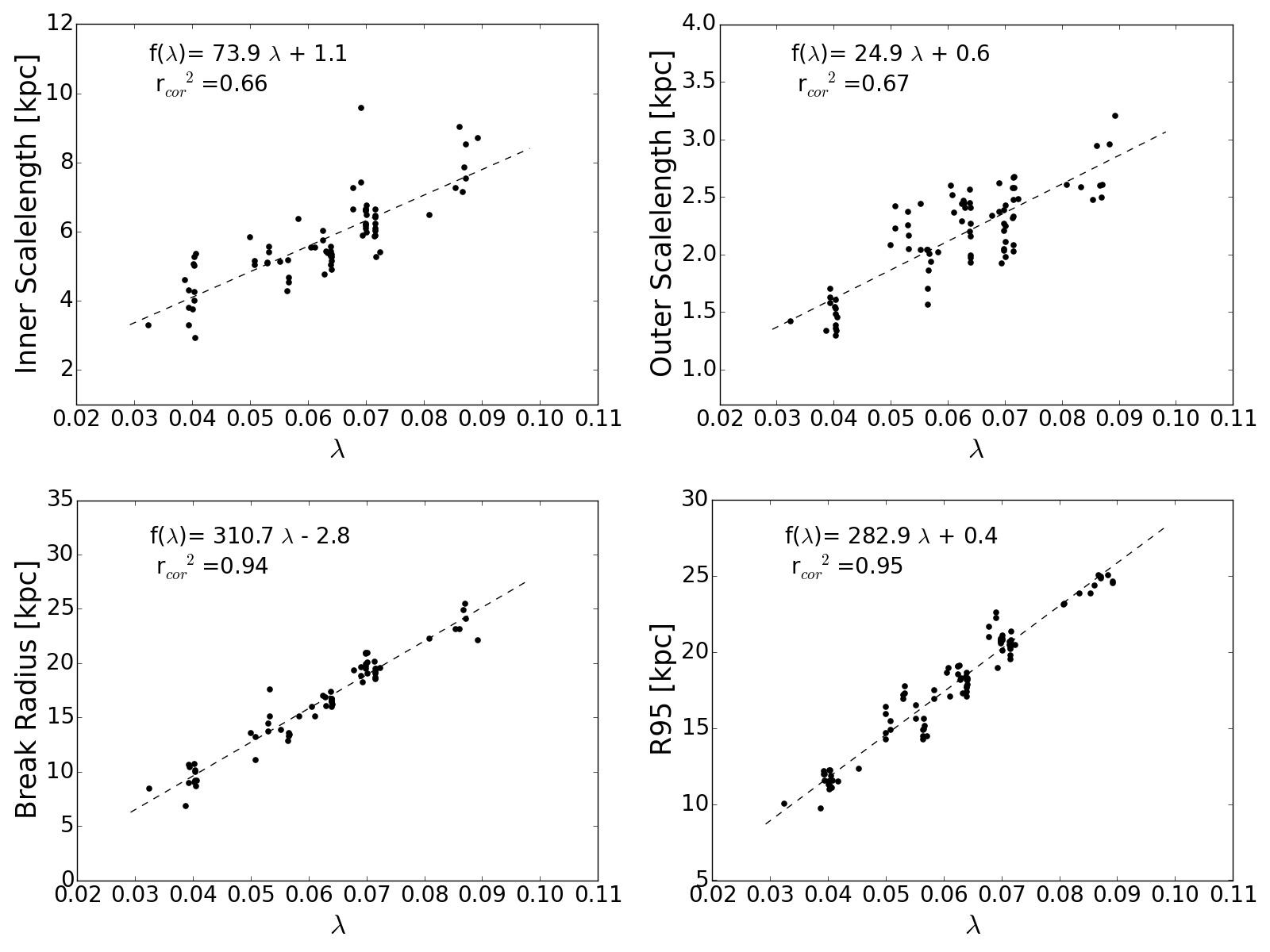}
% \vspace*{-1.0 cm}
 \caption{Characteristic radii of type II profiles from a number of
   simulations. From left to right and from top to bottom we have the
   inner disc scalelength, the outer one, the break radius and the
   spherical radius
   containing 95\%  of the stellar mass. The
   corresponding correlation coefficients and the regression line
   equations
   are given in the upper left corner of each panel. Each point
   corresponds to one simulation.}
   \label{fig:typeIIb}
\end{center}
\end{figure}

An interesting question in this context is the relation between the
final stellar density distribution in the galaxy and the total angular
momentum of the system. A number of works (e.g. \cite{Dalcanton.SS.97},
\cite{KimLee.13}, \cite{Herpich.SD.15}) examined this question in the framework of
isolated disc galaxy formation. P16 extended this to the scenario
proposed in A16a. In this case,  
except for the internal angular momentum of the two protogalaxies, one
should take into account also 
the orbital angular momentum and it is not clear how this
will influence the final stellar density distribution and what the global
effect of the merging on this distribution will be.
The simulation galaxies used in P16 
have a few properties in common, namely they have
the same baryonic to total mass ratio and the
same initial mass distributions in the dark matter as well as in the gaseous
haloes. Furthermore, in order to be able to make comparisons between
galaxies in the same evolutionary stage, P16
considered all remnants roughly 8 Gyr after the merging started. 
They include, however, different protogalaxy mass ratios
(varying between 1:1 and 1:3), a number of different orbits of the
protogalaxies, different halo spin parameters, different
orientations of the discs before the merging, and even different
numerical modellings, e.g. with or without AGN feedback.

P16 also calculated for these remnants
the normalised values of the angular momentum defined as
$\lambda= J |E|^{0.5} / (G M^{2.5})$.
Here $E$, $J$ and $M$ are the total energy, angular momentum and mass of the
system, respectively, and $G$ is the gravitational constant. Fig.~\ref{fig:typeIIb}
shows four characteristic radii of the density distribution as a
function of $\lambda$. These are the inner and the outer disc
scalelengths, the break radius, and $R_{95}$, the spherical radius containing 95\%
of the total stellar mass. This figure includes all simulations for which the
corresponding fit to the disc radial density profile 
was considered satisfactory. As found in P16, there is
a clear correlation in all panels, showing that a simulation with
larger angular momentum will result in more extended discs (inner and
outer). Qualitatively, this means that, for a given initial baryonic 
mass, the extent of 
the disc galaxy formed from the remnant will be larger when the
angular momentum is larger, as expected. 

Note that the quantity most affected by the angular momentum
is the break radius, while the least affected is the outer
disc scale length. Indeed, as can be seen 
from the slopes of the four regression lines in
Fig.~\ref{fig:typeIIb}, for a
given change in $\lambda$, the change of the break radius is
roughly 12 times more than the corresponding change for the 
outer disc scale length. Note also that the correlations 
with the break radius and with $R_{95}$ are considerably tighter than
the ones with the disc scalelengths. %This could be due to the
%facts that the estimate of $R_{95}$ does not rely on any fit and that
%the break radius is more accurately determined from the fits than the inner
%and outer disc scale lengths, as P16 were able to determine by
%multiple trials. This would imply that the spread around the
%xregression line may give information on the quality of the fits.

\acknowledgements{I thank the organisers for inviting me to review
  dynamical models of the 
outskirts of disc galaxies in this interesting and inspiring
meeting. This work was supported by the DAGAL network from the People Programme (Marie Curie Actions) of the European Union's Seventh Framework Programme FP7/2007-2013/ under REA grant agreement PITN-GA-2011-289313, and by the CNES (Centre National
dude's Spatiales - France). I also thank my collaborators A. Bosma,
S. Rodionov,  
N. Peschken and J.-C. Lambert for many useful discussions, help with
technical aspects and encouragement. 
This work was granted access to the HPC resources of
[TGCC/CINES/IDRIS] under the allocation 2016047665 made by GENCI. 
It was also granted access to the HPC resources of Aix-Marseille
Universit\'e financed by the project Equip@Meso (ANR-10-EQPX-29-01) of
the program ``Investissements d Avenir" supervised by the Agence Nationale de la Recherche.}

%\section{Appendix}
%\appendix
\vspace{0.5cm}
\noindent
{\bf Appendix}
\\

\indent
The simulations discussed in Sect.~\ref{sec:outer-discs} start off 
with two protogalaxies composed each of a slowly rotating sphere of dark matter
and hot gas (A16). The orbits of these protogalaxies are such as to lead to a 
merging. The gas in each protogalaxy starts
cooling from the beginning of the simulation (i.e. well before the
merging) and falls towards the equatorial plane of its protogalaxy,
starting to form a disc component there. The
properties of these protogalaxies before the merging occurs
are nearer to those of galaxies at intermediate redshifts than those
of nearby spirals, as they should (A16). Indeed they are smaller, more gas rich, more
lumpy and less dynamically relaxed than the latter. During the
collision the two protogalaxy haloes merge into a single one, while the stars 
born in the protogalaxies undergo violent relaxation and form a classical
bulge. Stars born during the collision or very shortly after it  
undergo more like a strong shuffling and a large fraction of them
start forming the 
thick disc component. After the merging is completed, the thin disc starts
forming from the gas accreting from the now common halo (A16). 

Thus the classical bulge is made of the oldest stars
and has kinematics as those of the observed classical bulges,
the thin disc consists of the youngest stars and shows strong rotation while
the thick disc has an intermediate age population and rotates
slower than the thin disc.  
A16 find that this simple model does
surprisingly well when compared to observations. Comparisons include the
stellar and gaseous projected surface density profiles, the rotation
curve, the properties of the thick disc, the differences between the kinematics
of the classical bulge and those of the thin disc etc. It is possible
in this model to form even late type disc galaxies, with a ratio of classical
bulge to total stellar mass of around 10\%. A bar forms and
has a very realistic morphology, including a boxy/peanut bulge, ansae
and a barlens. A 
discy pseudobulge forms from the gas in the central region. In fact
the youngest stars are found in this component,  as well as in the
extended spirals in the disc.

\end{document}